\documentclass{article} % For LaTeX2e
\usepackage{nips12submit_e,times}
\usepackage[psamsfonts]{amssymb}
\usepackage[pdftex]{graphicx}
\usepackage{amssymb}
\usepackage{amsthm}

\def\1{\mathbf{1}}

\title{Efficient and optimal binary Hopfield associative \\ memory storage using minimum probability flow}

\author{
Christopher Hillar \\
Redwood Center for Theoretical Neuroscience \\
University of California, Berkeley \\
Berkeley, CA 94720 \\
\texttt{chillar@msri.org} \\
\And
Jascha Sohl-Dickstein \\
Redwood Center for Theoretical Neuroscience \\
University of California, Berkeley \\
Berkeley, CA 94720 \\
\texttt{jascha.sohldickstein@gmail.com} \\
\AND
Kilian Koepsell \\
Redwood Center for Theoretical Neuroscience \\
University of California, Berkeley \\
Berkeley, CA 94720 \\
\texttt{kilian@berkeley.edu} \\
}

\nipsfinalcopy % Uncomment for camera-ready version

\begin{document}

\maketitle

\begin{abstract}
We present an algorithm to store binary memories in a Hopfield neural network using minimum probability flow, a recent technique to fit parameters in energy-based probabilistic models.  In the case of memories without noise, our algorithm provably achieves optimal pattern storage (which we show is at least one pattern per neuron) and outperforms classical methods both in speed and memory recovery.  Moreover, when trained on noisy or corrupted versions of a fixed set of binary patterns, our algorithm finds networks which correctly store the originals.  We also demonstrate this finding visually with the unsupervised storage and clean-up of large binary fingerprint images from significantly corrupted samples.\footnote{Supported by an NSF All-Institutes Postdoctoral Fellowship administered by the Mathematical Sciences Research Institute through its core grant DMS-0441170 (CH) and by NSF grant IIS-0917342 (KK).  This work appeared at the 2012 Neural Information Processing Systems (NIPS) workshop on Discrete Optimization in Machine Learning (DISCML).}
\end{abstract}

\textbf{Introduction}.  In 1982, motivated by neural modeling work of \cite{little1974} and the Ising spin glass model from statistical physics \cite{ising25}, Hopfield introduced a method for the storage and retrieval of binary patterns in an auto-associative neural-network \cite{hopfield1982}.  Even today, this model and its various extensions \cite{cohen1983absolute, hinton1986learning} provide a plausible mechanism for memory formation in the brain.  However, existing techniques for training Hopfield networks suffer either from limited pattern capacity or excessive training time, and they exhibit poor performance when trained on unlabeled, corrupted memories.

Our main theoretical contributions here are the introduction of a tractable and neurally-plausible algorithm for the optimal storage of patterns in a Hopfield network, a proof that the capacity of such a network is at least one pattern per neuron, and a novel local learning rule for training neural networks.  Our approach is inspired by minimum probability flow \cite{sohl2011}, a recent technique for fitting probabilistic models that avoids computations with a partition function, the usually intractable normalization constant of a parameterized probability distribution.  

\begin{figure}[!t]
\centering
\includegraphics[width=4.0in]{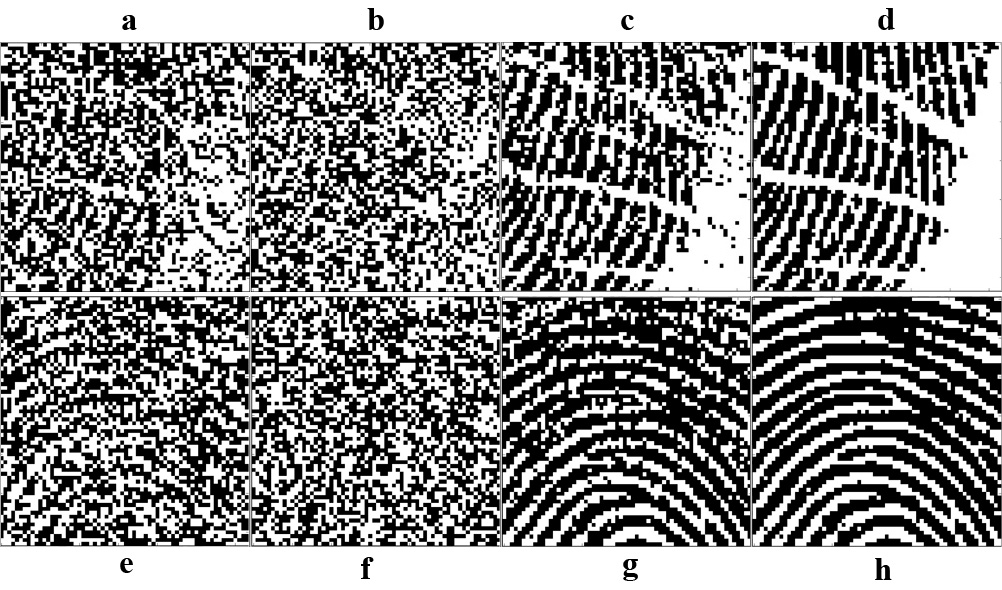}
\caption{\textbf{Learning from corrupted samples.} We stored $80$ fingerprints ($64 \times 64$ binary images) in a Hopfield network with $n = 64^2 = 4096$ nodes by minimizing the MPF objective (\ref{MPFobjective}) over a large set of randomly generated (and unlabeled) ``noisy" versions (each training pattern had a random subset of 1228 of its bits flipped; e.g., a,e).  After training, all $80$ fingerprints were stored as fixed-points of the network.  \textbf{a.} Sample training fingerprint with 30\% corruption. \textbf{b.} Fingerprint with 40\% corruption. \textbf{c.} State of  network after one dynamics update initialized at b.  \textbf{d.} Converged network dynamics equal to original fingerprint. \textbf{e-h.} As in {a-d}, but for different fingerprint.}
\label{fig7a}
\end{figure}

We also present several experimental results.  When compared with standard techniques for Hopfield pattern storage, our method is shown to be superior in efficiency and generalization.  Another finding is that our algorithm can store many patterns in a Hopfield network from highly corrupted (unlabeled) samples of them.  This discovery is also corroborated visually by the storage of $64 \times 64$ binary images of human fingerprints from highly corrupted versions, as in Fig.~\ref{fig7a}.

\textbf{Background.} A \textit{Hopfield network} $\mathcal{H} = (\mathbf J,\theta)$ on $n$ nodes  $\{1,\ldots,n\}$ consists of a symmetric \textit{weight matrix} $\mathbf J = \mathbf J^{\top} \in \mathbb R^{n \times n}$  with zero diagonal and a \textit{threshold vector} $\theta = (\theta_1,\ldots,\theta_n)^{\top} \in \mathbb R^n$.  We do not allow any row $\mathbf J_i$ of the matrix $\mathbf J$ to be zero.  The possible \textit{states} of the network are all length $n$ binary strings $\{0,1\}^n$, which we represent as binary column vectors $\mathbf x = (x_1,\ldots,x_n)^{\top}$, each $x_i  \in \{0,1\}$ indicating the state $x_i$ of node $i$.  Given any state $\mathbf x = (x_1,\ldots,x_n)^{\top}$, an (asynchronous) \textit{dynamical update} of $\mathbf x$ consists of replacing $x_i$ in $\mathbf x$ (in consecutive order starting with $i = 1$) with the value
\begin{equation}\label{Hopdynamics}
x_i = H(\mathbf J_i \mathbf x - \theta_i).
\end{equation}
Here, $\mathbf J_i$ is the $i$th row of $\mathbf J$ and $H$ is the \textit{Heaviside function} given by $H(r) = 1$ if $r > 0$ and $H(r) = 0$ if $r \leq 0$.

The \textit{energy} $E_{\mathbf x}$ of a binary pattern $\mathbf x$ in a Hopfield network  is  defined to be 
\begin{equation}\label{hopfield_energy}
E_{\mathbf x}(\mathbf J, \theta) := -\frac{1}{2}\mathbf x^{\top} \mathbf J \mathbf x + \theta^{\top}\mathbf  x = - \sum_{i < j} x_i x_j J_{ij} + \sum_{i=1}^n \theta_i x_i,
\end{equation}
identical to the energy function for an Ising spin glass.  In fact, the dynamics of a Hopfield network can be seen as 0-temperature Gibbs sampling of this energy function.  A fundamental  property of Hopfield networks is that asynchronous dynamical updates do not increase the energy (\ref{hopfield_energy}).  
Thus, after a finite number of updates, each initial state $\mathbf x$ converges to a \textit{fixed-point} $\mathbf x^{*} = (x_1^*,\ldots,x_n^*)^{\top}$ of the dynamics; that is, $x^{*}_i  = H(\mathbf J_i \mathbf x^{*} - \theta_i)$ for each $i$.

Given a binary pattern $\mathbf x$, the \textit{neighborhood} $\mathcal{N}(\mathbf x)$ of $\mathbf{x}$ consists of those binary vectors which are Hamming distance $1$ away from $\mathbf x$ (i.e., those with exactly one bit different from $\mathbf x$).   
We say that $\mathbf x$ is a \textit{strict local minimum} if every $\mathbf x' \in \mathcal{N}(\mathbf x)$ has a strictly larger energy:
\begin{equation}\label{energy_diff}
0 > E_{\mathbf x}  - E_{\mathbf x'} =  (\mathbf J_i \mathbf x - \theta_i)\delta_i,
\end{equation}
where $\delta_i = 1-2 x_i$ and $x_i$ is the bit that differs between $\mathbf x$ and $\mathbf x'$.  It is straightforward to verify that if $\mathbf x$ is a strict local minimum, then it is a fixed-point of the dynamics.

A basic problem is to construct Hopfield networks with a given set $\mathcal D$ of binary patterns as fixed-points or strict local minima of the energy function (\ref{hopfield_energy}). Such networks are useful for memory denoising and retrieval since corrupted versions of patterns in $\mathcal D$ will converge through the dynamics to the originals. Traditional approaches to this problem consist of iterating over $\mathcal D$ a \textit{learning rule} \cite{hertz1991} that updates a network's weights and thresholds given a training pattern $\mathbf x \in \mathcal D$.  For the purposes of this work, we call a rule \textit{local} when the learning updates to the three parameters $J_{ij}$, $\theta_{i}$, and $\theta_{j}$ can be computed with access solely to $x_i, x_j$, the feedforward inputs $\mathbf J_i \mathbf x$, $\mathbf J_j \mathbf x$, and the thresholds $\theta_{i}$, $\theta_{j}$; otherwise, we call the rule \textit{nonlocal}. The locality of a rule is an important feature in a network training algorithm because of its necessity in theoretical models of computation in neuroscience.
 
In \cite{hopfield1982}, Hopfield defined an \textit{outer-product learning rule} (OPR) for finding such networks.  OPR is a local rule since only the binary states of nodes $x_i$ and $x_j$ are required to update a coupling term $J_{ij}$ during training (and only the state of $x_i$ is required to update $\theta_i$).  Using OPR, at most $n/(4\log n)$ patterns can be stored without errors in an $n$-node Hopfield network \cite{weisbuch1985,mceliece1987}.  In particular, the ratio of patterns storable to the number of nodes using this rule is at most $1/(4\log n)$ memories per neuron, which approaches zero as $n$ increases.  If a small percentage of incorrect bits is tolerated, then approximately $0.15n$ patterns can be stored \cite{hopfield1982, amit1987}.

\begin{figure}[!t]
\centering
\includegraphics[width=5.3in]{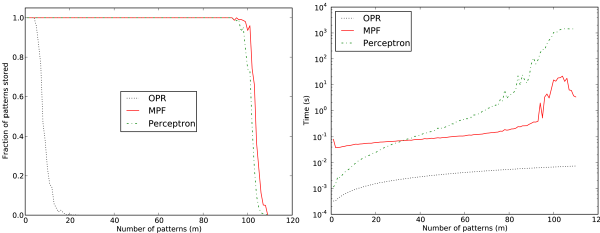}
\caption{({\bf Left}) Shows fraction of patterns made fixed-points of a Hopfield network using OPR (outer-product rule), MPF (minimum probability flow), and PER (perceptron) as a function of the number of randomly generated training patterns $m$.  Here, $n = 64$ binary nodes and we have averaged over $t = 20$ trials.  The slight difference in performance between MPF and PER is due to the extraordinary number of iterations required for PER to achieve perfect storage of patterns near the critical pattern capacity of the Hopfield network.  See also Fig.~\ref{fig1}. ({\bf Right}) Shows time (on a log scale) to train a Hopfield network with $n = 64$ neurons to store $m$ patterns using OPR, PER, and MPF (averaged over $t=20$ trials).}
\vspace{-.25cm}
\label{fig1}
\end{figure}

\begin{figure}[!t]
\centering
\includegraphics[width=5.4in]{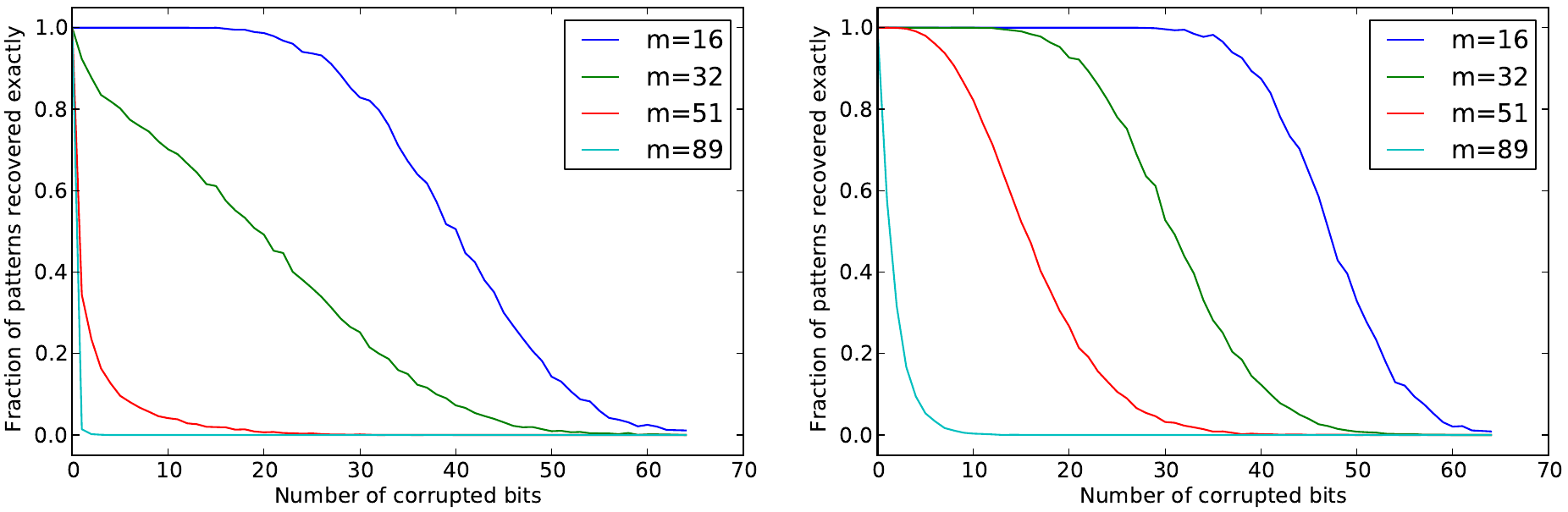}
\caption{Shows fraction of exact pattern recovery for a perfectly trained $n = 128$ Hopfield network using rules PER (figure on the left) and MPF (figure on the right) as a function of bit corruption at start of recovery dynamics for various numbers $m$ of patterns to store.  We remark that this figure and the next do not include OPR as its performance was far worse than MPF or PER.}
\vspace{-.25cm}
\label{fig2}
\end{figure}

\begin{figure}[!t]
\centering
\includegraphics[width=3.0in]{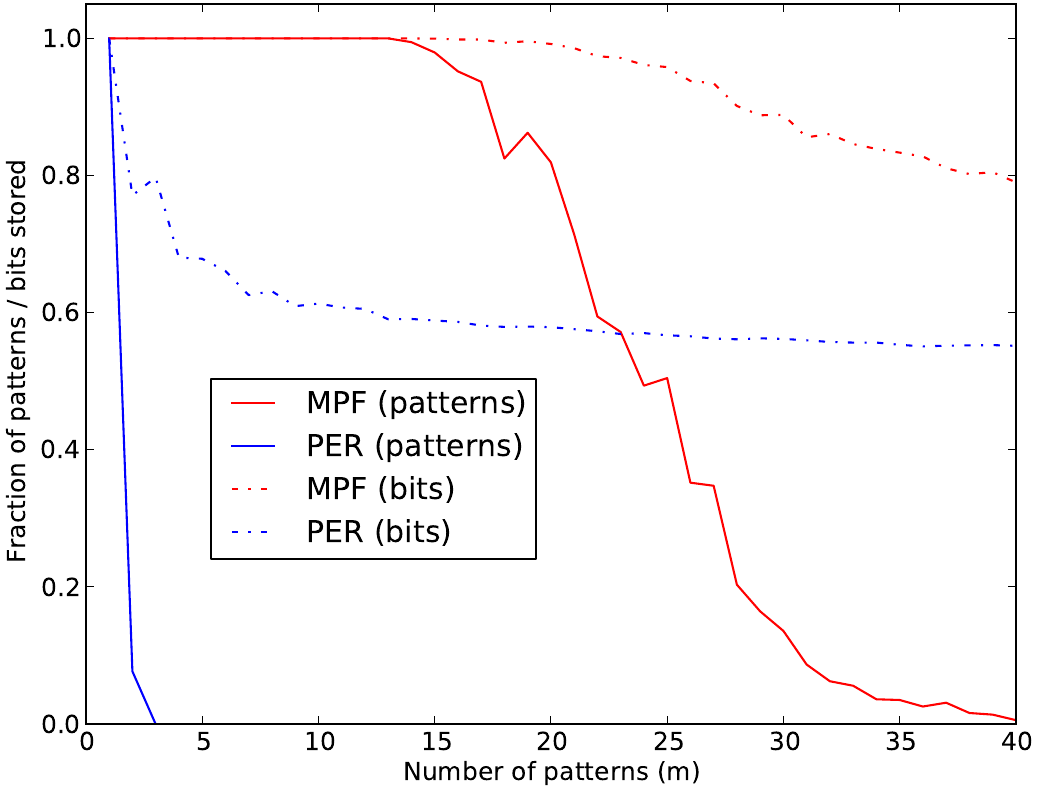}
\caption{Shows fraction of patterns (shown in red for MPF and blue for PER) and fraction of bits (shown in dotted red for MPF and dotted blue for PER) recalled of trained networks (with $n = 64$ nodes each) as a function of the number of patterns $m$ to be stored.  Training patterns were presented repeatedly with 20 bit corruption (i.e., 31\% of the bits flipped). (averaged over t = 13 trials.)}
\vspace{-.25cm}
\label{fig5a}
\end{figure}

The \textit{perceptron learning rule} (PER) \cite{rosenblatt1957perceptron, minsky1988, hertz1991} provides an alternative method to store patterns in a Hopfield network \cite{jinwen1993}.  PER is also a local rule since updating $J_{ij}$ requires only $\mathbf J_i\mathbf x$ and $\mathbf J_j\mathbf x$ (and updating $\theta_i$ requires $\mathbf J_i\mathbf x$).  Unlike OPR, it achieves optimal storage capacity, in that if it is possible for a collection of patterns $\mathcal D$ to be fixed-points of a Hopfield network, then PER will converge to parameters $\mathbf J, \theta$ for which all of $\mathcal D$ are fixed-points.  However, training frequently takes many parameter update steps (see Fig.~\ref{fig1}), and the resulting Hopfield networks do not generalize well (see Fig.~\ref{fig2}) nor store patterns from corrupted samples (see Fig.~\ref{fig5a}).

Despite the connection to the Ising model energy function, and the common usage of Ising spin glasses (otherwise referred to as Boltzmann machines \cite{hinton1986learning}) to build probabilistic models of binary data, we are aware of no previous work on associative memories that takes advantage of a probabilistic interpretation during training.  (Although probabilistic interpretations have been used for pattern recovery \cite{Sommer1998}.)

\textbf{Theoretical Results.}  We give an efficient algorithm for storing at least $n$ binary patterns as strict local minima (and thus fixed-points) in an $n$-node Hopfield network, and we prove that this algorithm achieves the optimal storage capacity achievable in such a network.  We also present a novel local learning rule for the training of neural networks.

Consider a collection of $m$ binary $n$-bit patterns $\mathcal{D}$ to be stored as strict local minima in a Hopfield network.  Not all collections of $m$ such patterns $\mathcal{D}$ can so be stored; for instance, from (\ref{energy_diff}) we see that no two binary patterns one bit apart can be stored simultaneously.  Nevertheless, we say that the collection $\mathcal{D}$ \textit{can be stored as local minima} of a Hopfield network if there is some $\mathcal{H} = (\mathbf J, \theta)$ such that each $\mathbf x \in \mathcal{D}$ is a strict local minimum of the energy function $E_{\mathbf x}(\mathbf J,\theta)$ in (\ref{hopfield_energy}).

The \textit{minimum probability flow} (MPF) objective function \cite{sohl2011} given the collection $\mathcal{D}$ is
\begin{equation}\label{MPFobjective}
K_{\mathcal{D}}(\mathbf J,\theta) := \sum_{\mathbf x \in \mathcal{D}} \ \sum_{\mathbf x' \in \mathcal{N}(\mathbf x)} \exp \left(\frac{E_{\mathbf x}-E_{\mathbf x'}}{2}\right).
\end{equation}
The function in (\ref{MPFobjective}) is infinitely differentiable and (generically, strictly) convex in the parameters.
Notice that when $K_{\mathcal{D}}(\mathbf J,\theta)$ is small, the energy differences $E_{\mathbf x}-E_{\mathbf x'}$ between $\mathbf x \in \mathcal{D}$ and patterns $\mathbf x'$ in neighborhoods $\mathcal{N}(\mathbf x)$ will satisfy (\ref{energy_diff}), making $\mathbf x$ a fixed-point of the dynamics.

As the following result explains, minimizing (\ref{MPFobjective}) given a storable set of patterns will determine a Hopfield network storing those patterns.

\textbf{Theorem 1.} \textit{If a set of binary vectors $\mathcal{D}$ can be stored as local minima of a Hopfield network, then minimizing the convex MPF objective (\ref{MPFobjective}) will find such a network.}

Our next result is that at least $n$ patterns in an $n$-node Hopfield network can be stored by minimizing (\ref{MPFobjective}) (see also \cite{personnaz1986, kanter1987}).  To make this statement mathematically precise, we introduce some notation.  Let $r(m,n) < 1$ be the probability that a collection of $m$ binary patterns chosen uniformly at random from all ${2^n \choose m}$ $m$-element subsets of $\{0,1\}^n$ can be made local minima of a Hopfield network.  The \textit{pattern capacity} (per neuron) of the Hopfield network is defined to be the supremum of all real numbers $a > 0$ such that $\lim_{n \to \infty} r(an,n) = 1$.

\textbf{Theorem 2.} \textit{The capacity of an $n$-node Hopfield network is at least $1$ pattern per neuron.}

In other words, for any fixed $a < 1$, the fraction of all subsets of $m = an$ patterns that can be made strict local minima (and thus fixed-points) of a Hopfield network with $n$ nodes converges to $1$ as $n$ tends to infinity.  Moreover, by Theorem 1, such networks can be found by minimizing (\ref{MPFobjective}). Experimental evidence suggests that the capacity limit above is $1$ for all $a < 1.5$, but converges to $0$ for $a > 1.7$ (see Fig.~\ref{fig1}). Although the Cover bound \cite{cover1965} forces $a \leq 2$, it is an open problem to determine the exact critical value of $a$ (i.e., the exact pattern capacity of the Hopfield network).  

We close this section by defining a new learning rule for a neural network. In words, the \textit{minimum probability flow learning rule} (MPF) takes an input training pattern $\mathbf x$ and moves the parameters $(\mathbf J,\theta)$ a small amount in the  direction of steepest descent of the MPF objective function $K_{\mathcal{D}}(\mathbf J,\theta)$ with $\mathcal{D} = \{\mathbf x\}$.  These updates for $J_{ij}$ and $\theta_i$ take the form (where again, \mbox{$\mathbf \delta = \mathbf 1 - 2\mathbf x$}):
\begin{eqnarray}\label{online_MPF_update}
\Delta J_{ij} &\propto & -\delta_i x_j e^{\frac{1}{2} \left( \mathbf J_i \mathbf x - \theta_i \right)\delta_i } 
	 - \delta_j x_i e^{\frac{1}{2} \left( \mathbf J_j \mathbf x - \theta_j \right)\delta_j} \\
\Delta \theta_i &\propto &\delta_i e^{ \frac{1}{2} \left( \mathbf J_i \mathbf x - \theta_i \right)\delta_i}. \label{online_MPF_thresh_update}
\end{eqnarray}
It is clear from (\ref{online_MPF_update}) and (\ref{online_MPF_thresh_update}) that MPF is a local learning rule.

\textbf{Experimental results.}  We performed several experiments comparing standard techniques for fitting Hopfield networks with  minimizing the MPF objective function (\ref{MPFobjective}).  All computations were performed on standard desktop computers, and we used used the limited-memory Broyden-Fletcher-Goldfarb-Shanno (L-BFGS) algorithm \cite{nocedal1980} to minimize (\ref{MPFobjective}).

In our first experiment, we compared MPF to the two methods OPR and PER for finding $64$-node Hopfield networks storing a given set of patterns $\mathcal{D}$.  For each of $20$ trials, we used the three techniques to store a randomly generated set of $m$ binary patterns, where $m$ ranged from $1$ to $120$.  The results are displayed in Fig.~\ref{fig1} and support the conclusions of Theorem 1 and Theorem 2.  

To evaluate the efficiency of our method relative to standard techniques, we compared training time of a $64$-node network as in Fig.~\ref{fig1} with the three techniques OPR, MPF, and PER.  The resulting computation times are displayed in Fig.~\ref{fig1} on a logarithmic scale.  Notice that computation time for MPF and PER significantly increases near the capacity threshold of the Hopfield network.

For our third experiment, we compared the denoising performance of MPF and PER.  For each of four values for $m$ in a $128$-node Hopfield network, we determined weights and thresholds for storing all of a set of $m$ randomly generated binary patterns using both MPF and PER.  We then flipped $0$ to $64$ of the bits in the stored patterns and let the dynamics (\ref{Hopdynamics}) converge (with weights and thresholds given by MPF and PER), recording if the converged pattern was identical to the original pattern or not.  Our results are shown in Fig~\ref{fig2}, and they demonstrate the superior corrupted memory retrieval performance of MPF.

We also tested the how the efficiency of our algorithm scales with the number of nodes $n$.  For varying $n$, we fit $m = n/4$ (randomly generated) patterns in $n$-node Hopfield networks using MPF over $50$ trials and examined  the training time.  Our experiments show that the average training time to fit Hopfield networks with MPF is well-modeled by a polynomial $O(n^{5/2})$.

A surprising final finding in our investigation was that MPF can store patterns from highly corrupted or noisy versions on its own and without supervision.  This result is explained in Fig~\ref{fig5a}.  To illustrate the experiment visually, we stored $m = 80$ binary fingerprints in a $4096$-node Hopfield network using a large set of training samples which were corrupted by flipping at random $30\%$ of the original bits; see Fig.~\ref{fig7a} for more details.  

\textbf{Discussion.} We have presented a novel technique for the storage of patterns in a Hopfield associative memory.   The first step of the method is to fit an Ising model using minimum probability flow learning to a discrete distribution supported equally on a set of binary target patterns.  Next, we use the learned Ising model parameters to define a Hopfield network.  We show that when the set of target patterns is storable, these steps result in a Hopfield network that stores all of the patterns as fixed-points.  We have also demonstrated that the resulting (convex) algorithm outperforms current techniques for training Hopfield networks.  We have shown improved recovery of memories from noisy patterns  and improved training speed as compared to training by PER.  We have demonstrated optimal storage capacity in the noiseless case, outperforming OPR.  We have also demonstrated the unsupervised storage of memories from heavily corrupted training data.  Furthermore, the learning rule that results from our method is local; that is, updating the weights between two units requires only their states and feedforward input.   As MPF allows the fitting of large Hopfield networks quickly, new investigations into the structure of Hopfield networks are posssible \cite{HTK}.  It is our hope that the robustness and speed of this learning technique will enable practical use of Hopfield associative memories in computational neuroscience, computer science, and scientific modeling.

\bibliographystyle{IEEEtran}
\bibliography{exp_storage_hopfield}

\end{document}